\documentclass{ws-procs9x6}

\begin{document}

\newcommand {\bea}{\begin{eqnarray}}
\newcommand {\eea}{\end{eqnarray}}
\newcommand {\be}{\begin{equation}}
\newcommand {\ee}{\end{equation}}
\newcommand {\qslash}{q\!\!\!/}
\newcommand {\muslash}{\mu\!\!\!/}
\newcommand {\Dslash}{D\!\!\!/}
\newcommand {\Slash}{\!\!\!\!/\,}

\title{Effective Theory of Superfluid Quark Matter}

\author{Thomas Sch{\"a}fer}

\address{Department of Physics, North Carolina State University,\\
Raleigh, NC 27695\\
and Riken-BNL Research Center, Brookhaven National Laboratory,\\
Upton, NY 11973 }


\maketitle

\abstracts{We provide a brief introduction to the high density 
effective theory of QCD. As an application, we consider the 
instanton correction to the perturbatively generated gap
in the color superconducting phase. We show that the instanton 
correction becomes large for $\mu\sim 1.25$ GeV in $N_f=2$ QCD, 
and for $\mu\sim 750$ MeV in $N_f=3$ QCD with a massive strange 
quark. We also study some other numerical issues related 
to the magnitude of the gap. We find, in particular, that a 
renormalization group improved gap equation does not give
results that are substantially different from a gap equation
with a fixed coupling.}

\section{Introduction}
\label{sec_intro}

  Over the last several years we have seen rapid progress in the 
theoretical study of very dense hadronic matter. Many new phases 
of strongly interacting matter, such as color superconducting quark 
matter and color-flavor locked matter have been predicted 
\cite{Bailin:1984bm,Alford:1998zt,Rapp:1998zu,Alford:1999mk,Schafer:1999fe}.
Reviews of these developments can be found in
\cite{Rajagopal:2000wf,Alford:2001dt,Nardulli:2002ma,Schafer:2003vz,Rischke:2003mt}. 
Exotic phases of matter at high baryon density may be realized in 
nature in the cores of neutron stars. In order to study this possibility 
quantitatively we would like to develop a systematic framework that will 
allow us to determine the exact nature of the phase diagram as a function 
of the density, temperature, the quark masses, and the lepton chemical 
potentials, and to compute the low energy properties of these phases. In 
this contribution we give a brief review of an attempt to use effective 
field theory methods to address this problem.

\section{High Density Effective Theory}
\label{sec_hdet}

 At high baryon density the relevant degrees of freedom are 
particle and hole excitations which move with the Fermi 
velocity $v$. Since the momentum $p\sim v\mu$ is large, 
typical soft scatterings cannot change the momentum by very 
much. An effective field theory of particles and holes 
in QCD is given by \cite{Hong:2000tn,Hong:2000ru}
\be
\label{l_hdet}
L=\sum_{v}
 \psi_{v}^\dagger (iv\cdot D) \psi_{v} 
 -\frac{1}{4}G^a_{\mu\nu} G^a_{\mu\nu}+ \ldots ,
\ee
where $v_\mu=(1,\vec{v})$. The field describes particles and 
holes with momenta $p=\mu\vec{v}+l$, where $l\ll\mu$. We will 
write $l=l_0+l_{\|}+l_\perp$ with $\vec{l}_{\|}=\vec{v}(\vec{l}
\cdot \vec{v})$ and $\vec{l}_\perp = \vec{l}-\vec{l}_{\|}$. In 
order to take into account the entire Fermi surface we have to 
cover the Fermi surface with patches labeled by the local Fermi 
velocity. The number of these patches is $n_v\sim (\mu^2/
\Lambda_\perp^2)$ where $\Lambda_\perp \ll\mu$ is the cutoff on 
the transverse momenta $l_\perp$. 

 Higher order terms are suppressed by powers of $1/\mu$. As 
usual we have to consider all possible terms allowed by the 
symmetries of the underlying theory. At $O(1/\mu)$ we have
\be
L= \sum_{v}\left\{
  -\frac{1}{2\mu} \psi_{v}^\dagger D_\perp^2 \psi_{v}
  - g  \psi_{v}^\dagger \frac{\sigma^{\mu\nu} G_{\mu\nu}^\perp}
   {4\mu}\psi_v \right\}. 
\ee
At higher order in $1/\mu$ there is an infinite tower of operators 
of the form $\mu^{-n}\psi^\dagger_v D_\perp^{2n_1}(\bar{v}
\cdot D)^{n_2} \psi_v$ with $\bar{v}=(1,-\vec{v})$ and $n=
2n_1+n_2-1$. At $O(1/\mu^2)$ the effective theory contains 
four-fermion operators \cite{Schafer:2003jn}
\bea 
L&=&  \frac{1}{\mu^2} \sum_{v_i} \sum_{\Gamma,\Gamma'} 
  c^{\Gamma\Gamma'}(\vec{v}_1\cdot\vec{v}_2,\vec{v}_1\cdot\vec{v}_3,
   \vec{v}_2\cdot\vec{v}_3) \nonumber \\
  & & \mbox{}\hspace{0.5cm}\cdot 
   \Big(\psi_{v_1} \Gamma \psi_{v_2}\Big)
   \Big(\psi^\dagger_{v_3}\Gamma'\psi^\dagger_{v_4}\Big)
   \delta(v_1+v_2-v_3-v_4).
\eea
There are two types of operators that are compatible with the 
restriction $v_1+v_2=v_3+v_4$. The first possibility is that 
both the incoming and outgoing fermion momenta are back-to-back. 
This corresponds to the BCS interaction
\be
\label{c_bcs}
L=  \frac{1}{\mu^2}\sum_{v,v'}\sum_{\Gamma,\Gamma'}
  V_l^{\Gamma\Gamma'} R_l^{\Gamma\Gamma'}(\vec{v}\cdot\vec{v}') 
    \Big(\psi_{v} \Gamma \psi_{-v}\Big)
   \Big(\psi^\dagger_{v'}\Gamma'\psi^\dagger_{-v'}\Big),
\ee
where $\vec{v}\cdot\vec{v}'=\cos\theta$ is the scattering 
angle and $R_l^{\Gamma\Gamma'}(\vec{v}\cdot\vec{v}')$ is a 
set of orthogonal polynomials. The second possibility is 
that the final momenta are equal to the initial momenta up 
to a rotation around the axis defined by the sum of the incoming 
momenta. The relevant four-fermion operator is 
\be
\label{c_flp}
L=  \frac{1}{\mu^2}\sum_{v,v',\phi}\sum_{\Gamma,\Gamma'}
  F_l^{\Gamma\Gamma'}(\phi) R_l^{\Gamma\Gamma'}(\vec{v}\cdot\vec{v}') 
    \Big(\psi_{v} \Gamma \psi_{v'}\Big)
   \Big(\psi^\dagger_{\tilde{v}}\Gamma'\psi^\dagger_{\tilde{v}'}
  \Big),
\ee
where $\tilde{v},\tilde{v}'$ are the vectors obtained from 
$v,v'$ by a rotation around $v_{tot}=v+v'$ by the angle $\phi$.

 The four-fermion operators in the effective theory can 
be determined by matching moments of quark-quark scattering 
amplitudes between QCD and the effective theory. The matching 
conditions involve on-shell scattering amplitudes in BCS and 
forward kinematics. The scattering amplitude in the effective theory 
contains almost collinear gluon exchanges which do not change 
the velocity label of the quarks as well as four-fermion 
operators which correspond to scattering involving different 
patches on the Fermi surface. There are several contributions
in the microscopic theory that are absorbed into four-fermion
operators in the effective theory. Examples are large angle
scatterings \cite{Schafer:2003jn} and non-perturbative 
instanton effects \cite{Schafer:2002ty}.

\section{Power Counting}
\label{sec_pow}

 In this section we briefly discuss the power counting in the 
high density effective theory. We will denote the small scale
by $l\ll\mu$. We first discuss the scaling properties of a 
generic operator. We assume that $v\cdot D$ scales as $l$, 
$\psi_v$ scales as $l^{3/2}$, $A_\mu$ scales as $l$, and 
$\vec{D}_\perp,\bar{v}\cdot D\sim l$. Complication arise 
because not all loop diagrams scale as $l^4$. In fermion 
loops sums over patches and integrals over transverse momenta 
can combine to give integrals that are proportional to the 
surface area of the Fermi sphere, 
\be
\label{hard_int}
 \frac{1}{2\pi}\sum_{\vec{v}}\int\frac{d^2l_\perp}{(2\pi)^2}
 =\frac{\mu^2}{2\pi^2}\int \frac{d\Omega}{4\pi}.
\ee
\begin{figure}
\begin{center}
\leavevmode
\includegraphics[width=5cm,clip=true]{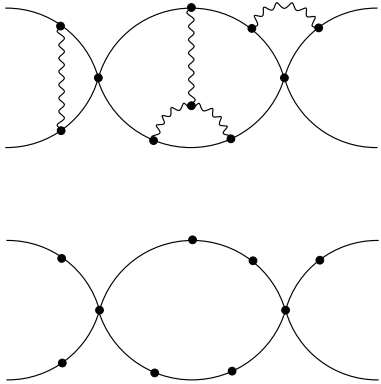}
\end{center}
\caption{Counting hard loops in the effective field theory.
If all (soft) gluon lines are removed the remaining fermionic
loops contain sums over the velocity index.}
\label{fig_lcount}
\end{figure}
These loop integrals scale as $l^2$, not $l^4$. In the 
following we will refer to loops that scale as $l^2$ as 
``hard loops'' and loops that scale as $l^4$ as ``soft 
loops''. In order to take this distinction into account we 
define $V_k^S$ and $V_k^H$ to be the number of soft and hard 
vertices of scaling dimension $k$. A vertex is called soft if it 
contains no fermion lines. In order to determine the $l$ 
counting of a general diagram in the effective theory we 
remove all gluon lines from the graph, see Fig.~\ref{fig_lcount}. 
We denote the number of connected pieces of the remaining graph 
by $N_C$. Using Euler identities for both the initial and the 
reduced graph we find that the diagram scales as $l^\delta$ with 
\be
\label{pc_imp}
 \delta = \sum_i \left[ (k-4)V_k^S + (k-2-f_k)V_k^H\right]
 +E_Q +4 - 2N_C.
\ee
Here, $f_k$ denotes the number of fermion fields in  
a hard vertex, and $E_Q$ is the number of external quark 
lines. We observe that in general the scaling dimension 
$\delta$ increases with the number of higher order vertices, 
but there are two important exceptions. 

 First we observe that the number of disconnected fermion
loops, $N_C$, reduces the power $\delta$. Each disconnected
loop contains at least one power of the coupling constant, 
$g$, for every soft vertex. As a result, fermion loop insertions 
in gluon $n$-point functions spoil the power counting if the gluon 
momenta satisfy $l\sim g\mu$. This implies that for $l<g\mu$ the 
high density effective theory becomes non-perturbative and
fermion loops in gluon $n$-point functions have to be resummed. 
The generating functional for hard dense loops in gluon $n$-point 
functions is given by \cite{Braaten:1991gm,Braaten:1992jj}
\be 
\label{S_hdl}
L_{HDL} = -m^2\int\frac{d\Omega}{4\pi}
 {\rm Tr}\,G_{\mu \alpha} 
  \frac{\hat{P}^\alpha \hat{P}^\beta}{(\hat{P}\cdot D)^2} 
G^\mu_{\,\beta},
\ee
where $m^2=N_fg^2\mu^2/(4\pi^2)$ and the angular integral 
corresponds to an average over the direction of $\hat{P}_\alpha 
=(1,\hat{p})$. For momenta $l<g\mu$ we have to add $L_{HDL}$ to 
$L_{HDET}$. In order not to over-count diagrams we have to remove 
at the same time all diagrams that become disconnected if all soft 
gluon lines are deleted. Note that the high density effective 
theory is different from the standard hard dense loop approximation, 
because hard loops are resummed only in gluon Green functions,
but not in quark Green functions or quark-gluon vertex functions. 

 The second observation is that the power counting for hard vertices 
is modified by a factor that counts the number of fermion lines in the 
vertex. It is easy to see that four-fermion operators without extra 
derivatives are leading order ($k-2-f_k=0$), but terms with more than 
four fermion fields, or extra derivatives, are suppressed. This result 
is familiar from the effective field theory analysis of theories with 
short range interactions \cite{Shankar:1993pf,Polchinski:1992ed}.

\section{Color Superconductivity}
\label{sec_gap}

 In the last section we saw that hard loops lead to non-perturbative 
effects in the effective theory that require resummation at the scale 
$l\sim g\mu$. In addition to that, there are logarithmic divergences that
have to be resummed at exponentially small scales $l\sim \mu\exp(-c/g)$.
The most important effect of this type is the BCS instability in the
quark-quark scattering amplitude. This instability leads to the 
formation of a gap in the single particle spectrum. We can take this
effect into account in the high density effective theory 
by including a tree level gap term 
\be 
\label{gap}
L= \Delta\, R_l^\Gamma(\vec{v}\cdot \hat\Delta)
  \psi_{-v}\sigma_2\Gamma \psi_{v} + h.c..
\ee
The Dirac matrix $\Gamma$ and the angular factor $R_l^\Gamma(x)$ 
determine the helicity and partial wave channel. The magnitude of 
the gap is determined variationally, by requiring the free energy 
to be stationary order by order in perturbation theory. 

 At leading order in the high density effective theory the 
variational principle for the gap $\Delta$ gives the Dyson-Schwinger 
equation 
\be 
\label{gap_1}
\Delta(p_4) =\frac{2g^2}{3} \int \frac{d^4q}{(2\pi)^4} 
 \frac{\Delta(q_4)}{q_4^2+l_q^2+\Delta(q_4)^2}
  v_\mu v_\nu D_{\mu\nu}(p-q) ,
\ee
where we have restricted ourselves to angular momentum zero and 
the color anti-symmetric $[\bar{3}]$ channel. $D_{\mu\nu}$ is the 
hard dense loop resummed gluon propagator. We also note that 
equ.~(\ref{gap_1}) only contains collinear exchanges. According 
to the arguments give in Sect.~\ref{sec_pow} four-fermion operators 
are of leading order in the HDET power counting. However, even 
though collinear exchanges and four-fermion operators have the 
same power of $l$, collinear exchanges are enhanced by a logarithm 
of the small scale. As a consequence, we can treat four-fermion
operators as a perturbation. 

  Sine the electric interaction is screened it is possible to absorb 
electric gluon exchanges into four-fermion operators. At leading 
order in the high density theory the gap equation is completely
determined by the collinear divergence in the magnetic gluon 
exchange interaction. This IR divergence is independent 
of the helicity and angular momentum channel. We have 
\be 
\label{gap_2}
\Delta(p_4) =\frac{g^2}{18\pi^2}
\int_0^{\Lambda_{\|}}  dq_4
\frac{\Delta(q_4)}{\sqrt{q_4^2+\Delta(q_4)^2}}
\log\left( \frac{\Lambda_\perp}{|p_4^2-q_4^2|^{1/2}}\right).
\ee
The leading logarithmic behavior is independent of the ratio 
of the cutoffs and we can set $\Lambda_{\|}=\Lambda_\perp
=\Lambda$. We introduce the dimensionless variables 
variables $x=\log(2\Lambda/(q_4+\epsilon_q))$ and $y
=\log(2\Lambda/(p_4+\epsilon_p)$ where $\epsilon_q=(q_4^2
+\Delta(q_4))^{1/2}$. In terms of dimensionless variables 
the gap equation is given by
\be 
\label{gap_3}
\Delta(y) = \frac{g^2}{18\pi^2}\int_0^{x_0} dx\, \Delta(x) K(x,y),
\ee
where $x_0=\log(2\Lambda/\Delta_0)$ and $K(x,y)$ is the kernel 
of the integral equation. At leading order we can use the 
approximation $K(x,y)=\min(x,y)$. We can perform an additional 
rescaling $x=x_0\bar{x}$, $y=x_0\bar{y}$. Since the leading order 
kernel is homogeneous in $x$ and $y$ we can write the gap equation 
as an eigenvalue equation 
\be 
\label{gap_4}
\Delta(\bar{y}) = x_0^2 \frac{g^2}{18\pi^2}\int_0^1 d\bar{x}\,
\Delta(\bar{x}) K(\bar{x},\bar{y}),
\ee
where the gap function is subject to the boundary conditions $\Delta
(0)=0$ and $\Delta'(1)=0$. This integral equation has the solutions 
\cite{Son:1999uk} 
\be 
\label{gap_6}
\Delta_n(\bar{x}) = \Delta_{n,0} 
\sin\left( \frac{g}{3\sqrt{2}\pi}x_{0,n}\bar{x} \right), 
\hspace{0.2cm}
x_{0,n}= (2n+1)\frac{3\pi^2}{\sqrt{2}g}.
\ee
The physical solution corresponds to $n=0$ which gives the 
largest gap, $\Delta_0=2\Lambda \exp(-3\pi^2/(\sqrt{2}g))$.
Solutions with $n\neq 0$ have smaller gaps and are not 
global minima of the free energy.

\section{Higher Order Corrections to the Gap}
\label{sec_cor}

  The high density effective field theory enables us to perform a 
systematic expansion of the kernel of the gap equation in powers of 
the small scale and the coupling constant. It is not so obvious, 
however, how to solve the gap equation for more complicated kernels, 
and how the perturbative expansion of the kernel is related to the 
expansion of the solution of the gap equation. 

\begin{figure}
\begin{center}
\includegraphics[width=3cm]{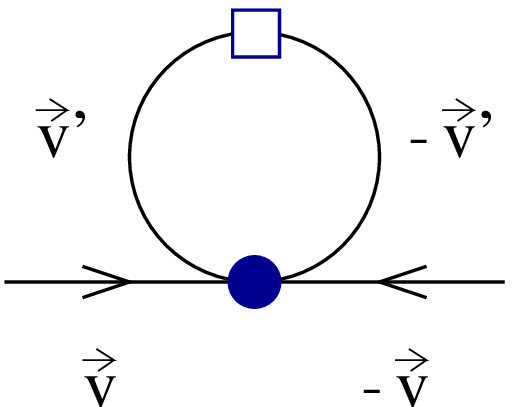}
\end{center}
\vspace*{0.5cm}
\begin{center}
\includegraphics[width=8cm]{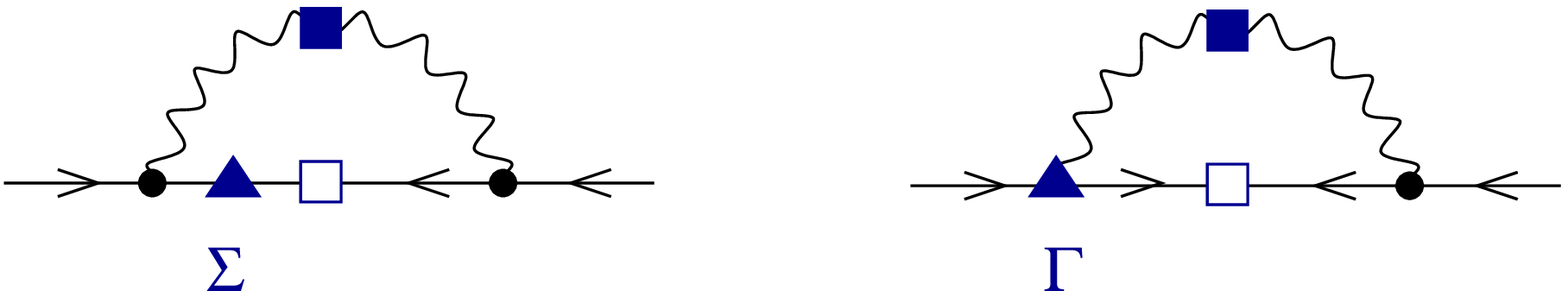}
\end{center}
\caption{Higher order corrections to the gap equation in the
high density effective theory. The diagrams shown in this 
figure correspond to four-fermion operators, fermion self-energy
corrections, and vertex corrections.}
\label{fig_sc2}
\end{figure}

 For this purpose it is useful to develop a perturbative method 
for solving the gap equation \cite{Brown:1999aq,Schafer:2003jn}.
We can write the kernel of the gap equation as $K(x,y)=K_0(x,y)
+\delta K(x,y)$, where $K_0(x,y)$ contains the leading IR 
divergence and $\delta K(x,y)$ is a perturbation. We expand
both the gap function $\Delta(x)$ and the eigenvalue $x_0$
order by order $\delta K$,
\bea
\Delta(\bar{x}) &=& \Delta^{(0)}(\bar{x}) + \Delta^{(1)}(\bar{x}) 
  +  \Delta^{(2)}(\bar{x}) + \ldots , \\[0.2cm]
\bar{x}_0 &=& \bar{x}_0^{(0)} + \bar{x}_0^{(1)} + \bar{x}_0^{(2)} 
  + \ldots ,
\eea
where we have defined $\bar{x}_0^2=g^2x_0^2/(18\pi^2)$. The 
expansion coefficients can be found using the fact that the 
unperturbed solutions given in equ.~(\ref{gap_6}) form an orthogonal 
set of eigenfunctions of $K_0$. The resulting expressions for 
$\bar{x}_0^{(i)}$ and $\Delta^{(i)}(\bar{x})$ are very similar
to Rayleigh-Schroedinger perturbation theory. At first order
we have
\bea
\label{dx_1}
\bar{x}_0^{(1)} &=& -\frac{1}{2}\left( \bar{x}_0^{(0)}\right)^2
\int_0^1 d\bar{x}\int_0^1d\bar{y}\, 
\Delta_0^{(0)}(\bar{x}) 
  \delta \bar{K}(x_0\bar{x},x_0\bar{y}) \Delta_0^{(0)}(\bar{y}),
  \\
\label{dc_1}
c^{(1)}_k &=& 
 \frac{\bar{x}_0^{(0)} }{1-\left(\frac{1}{2k+1}\right)^2}   
 \int_0^1 d\bar{x} \int_0^1 d\bar{y}\,
 \Delta^{(0)}_0(\bar{x}) \delta \bar{K}(x_0\bar{x},x_0\bar{y}) 
 \Delta^{(0)}_k(\bar{y}),
\eea
with $\Delta^{(1)}(x)=\sum c_k^{(1)}\Delta^{(0)}_k(x)$
and $\delta\bar{K}=g/(3\sqrt{2}\pi)\delta K$.

 We can now study the role of various corrections to the 
kernel. The simplest contribution arises from four-fermion 
operators. We find 
\be
\label{pert_ct}
\delta K(x_0\bar{x},x_0\bar{y}) = \log(b), 
\hspace{1cm}
b = \frac{512\pi^4\mu}{g^5\Lambda}
     \left(\frac{2}{N_f}\right)^{\frac{5}{2}}.
\ee
This contribution does not change the shape of the gap 
function but it gives an $O(g)$ correction to the eigenvalue 
$\bar{x}_0$. This corresponds to a constant pre-exponential 
factor, $\Delta^{(1)}=b\Delta^{(0)}$. An important advantage 
of the effective field theory method is that this factor is 
manifestly independent of the choice of gauge. The gauge 
independence of the pre-exponential factor is related to the 
fact that this coefficient is determined by four-fermion 
operators in the effective theory, and that these operators 
are determined by on-shell matching conditions. 

  Another effect that contributes to the eigenvalue at $O(g)$ 
is the fermion self energy \cite{Brown:1999aq,Brown:1999yd}. 
A one-loop calculation in the high density effective theory gives
\cite{Vanderheyden:1996bw,Brown:2000eh,Manuel:2000mk,Manuel:2000nh,Boyanovsky:2000bc}
\be 
\label{sigma_3}
\Sigma(p_4) =  \frac{g^2C_F}{12\pi^2}p_4 \log\left(
  \frac{\Lambda}{p_4}\right).
\ee
The correction to the kernel of the gap equation is 
\be
\delta \bar{K}(x_0\bar{x},x_0\bar{y}) =
 -\frac{g^2}{9\pi^2} \left(\bar{x}_0\bar{x}\right)
   K_0(x_0\bar{x},x_0\bar{y}),
\ee
and the shift in the eigenvalue is given by
\be
\label{x1_sig}
\bar{x}^{(1)}_0 = -\frac{1}{2}
 \left( \bar{x}_0^{(0)} \right)^2
 \langle 0| \delta \bar{K} | 0\rangle 
 =\frac{4+\pi^2}{8}\frac{g}{3\sqrt{2}\pi}, 
\ee
where $\langle 0| \delta \bar{K} | 0\rangle $ denotes the 
matrix element of the kernel between unperturbed gap 
functions, see equ.~(\ref{dx_1}). At this order in $g$, 
there is no contribution from the quark-gluon vertex 
correction. Note that the quark self energy correction 
makes an $O(g)$ correction to the eigenvalue, even though it 
is an $O(g^2)$ correction to the kernel. This is related 
to the logarithmic divergence in the self energy. The 
perturbative expansion of $\bar{x}_0$ is of the form
\be
\label{expansion}
\bar{x}_0 \sim g\log(\Delta) 
 = O(g^0) + O(g\log(g)) + O(g) + \ldots .
\ee
Brown et al.~argued that equ.~(\ref{x1_sig}) completes the 
$O(g)$ term. At this order the spin zero gap in the 2SC phase 
of $N_f=2$ QCD is \cite{Brown:1999yd,Wang:2001aq,Schafer:2003jn} 
\be
\label{gap_og}
\Delta = 512\pi^4 \mu g^{-5}
 e^{-\frac{4+\pi^2}{8}}e^{-\frac{3\pi^2}{\sqrt{2}g}}.
\ee
In other spin or flavor channels the relevant four fermion
operators are different and the pre-exponential factor is modified 
\cite{Schafer:1999fe,Brown:1999yd,Schafer:2000tw,Schmitt:2002sc}. 
In the CFL phase of $N_f=3$ QCD the gap is suppressed by a 
factor $(2/3)^{5/2}2^{-1/3}$.

\section{Instanton Correction} 
\label{sec_ins}

\begin{figure}
\begin{center}
\leavevmode
\includegraphics[width=7cm,clip=true]{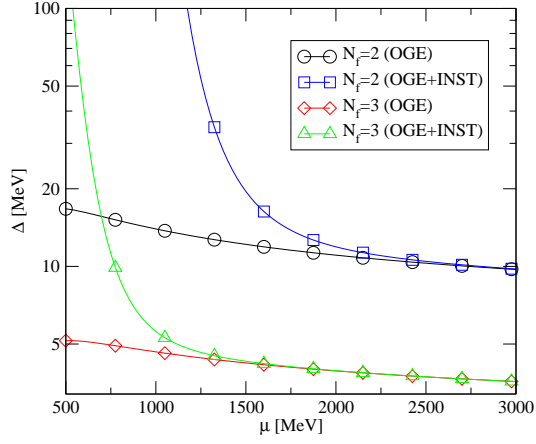}
\end{center}
\caption{Instanton correction to the perturbatively 
generated gap in QCD with $N_f=2$ and $N_f=2+1$ flavors.
We show the gap with and without instanton corrections.
In the $N_f=2+1$ case we have used $m_s=150$ MeV.}
\label{fig_gap_inst}
\end{figure}

 In this section we shall focus on the correction to the 
perturbatively generated gap parameter which is due to 
instantons. If the density is very large instanton effects
are exponentially small as compared to perturbative 
interactions. In this regime instantons are important 
for physical observables, like the mass of the $U(1)_A$
Goldstone mode, that receive no contribution from 
perturbative interactions, but they make no significant 
contribution to the gap. As the density is lowered
instanton effects grow. In the following we will 
use the methods described in the last section in order 
to determine the scale at which instanton contributions
to the gap become important. 

 Instantons induce a chirality changing four-fermion 
operator. In QCD with $N_f=3$ flavors and $m_{u,d}
\ll m_s$ we have
\bea
\label{l_nf2}
{ L} &=& \int n(\rho,\mu)d\rho\,
   \frac{2(2\pi\rho)^4\rho^3}{4(N_c^2-1)}
  m_s \epsilon_{f_1f_2}\epsilon_{g_1g_2} 
 \left( \frac{2N_c-1}{2N_c}
  (\bar\psi_{R,f_1} \psi_{L,g_1})
  (\bar\psi_{R,f_2} \psi_{L,g_2}) 
  \right. \nonumber \\
  & & \hspace{1cm}\mbox{}\left. 
   - \frac{1}{8N_c}
  (\bar\psi_{R,f_1} \sigma_{\mu\nu} \psi_{L,g_1})
  (\bar\psi_{R,f_2} \sigma_{\mu\nu} \psi_{L,g_2})
  + (L \leftrightarrow R ) \right),
\eea
where the sum over flavors runs over up and down 
quarks only, $f_{1,2}=g_{1,2}=(u,d)$, and the instanton 
size distribution $n(\rho,\mu)$ is given by
\bea
\label{G_I}
  n(\rho,\mu) &=& C_{N} \ \left(\frac{8\pi^2}{g^2}\right)^{2N_c}
 \rho^{-5}\exp\left[-\frac{8\pi^2}{g(\rho)^2}\right]
 \exp\left[-N_f\rho^2\mu^2\right],\\
 && C_{N} \;=\; \frac{0.466\exp(-1.679N_c)1.34^{N_f}}
    {(N_c-1)!(N_c-2)!}\, ,\\
 && \frac{8\pi^2}{g^2(\rho)} \;=\;
    -b\log(\rho\Lambda), \hspace{1cm}
    b = \frac{11}{3}N_c-\frac{2}{3}N_f \, .
\eea
Note that equ.~(\ref{l_nf2}) is an effective interaction 
for quarks near the Fermi surface. In particular, there 
are no form factors that have to be included. We also 
observe that the integration over sizes is cut off at 
$\rho\sim \mu^{-1}$. As a consequence, instanton effects
are of order $\exp(-8\pi^2/g^2(\mu))\sim (\Lambda_{QCD}
/\mu)^b$. In addition to the four-fermion vertex given 
in equ.~(\ref{l_nf2}) instantons also induce a six-fermion
operator. This operator has important physical effects, but
it does not contribute to the gap to leading order in the
effective interaction, so we will not consider it here. 
In QCD with $N_f=2$ massless flavors instantons induce
a four-fermion operator which can be obtained from 
equ.~(\ref{l_nf2}) by making the replacement $m_s\rho
\to 1$. 

 We can now compute the instanton correction to the 
kernel of the gap equation. In the three flavor case
we find $\delta K_I = \log(b_I)$ with $\log(b_I)=9G_I/g^2$ and
\bea
G_I(N_f\!=\!3) &=& \frac{C_N}{\mu^2}\left(\frac{m_s}{\mu}\right)
 \left(\frac{\Lambda_{QCD}}{\mu}\right)^{\beta_0}
  \left(\frac{8\pi^2}{g^2}\right)^{2N_c}
   \nonumber  \\
  & & \hspace{1.5cm}\mbox{}\cdot
 \frac{(2\pi)^2}{2(N_c^2-1)} 
  \frac{4(N_c+1)}{N_c}
  \frac{\Gamma\left(\frac{\beta_0+3}{2}\right)}
       {2N_f^{(\beta_0+3)/2}},
\eea
where $\beta_0=11N_c/3-2N_f/3$ is the first coefficient of the 
beta function. In the two flavor case we get 
\bea
G_I(N_f\!=\!2) &=& \frac{C_N}{\mu^2}
 \left(\frac{\Lambda_{QCD}}{\mu}\right)^{\beta_0}
  \left(\frac{8\pi^2}{g^2}\right)^{2N_c}
   \nonumber  \\
  & & \hspace{1.5cm}\mbox{}\cdot
 \frac{(2\pi)^2}{2(N_c^2-1)} 
  \frac{4(N_c+1)}{N_c}
  \frac{\Gamma\left(\frac{\beta_0}{2}+1\right)}
       {2N_f^{(\beta_0/2+1)}} . 
\eea
We note that the main difference is a suppression
factor $(m_s/\mu)$ in the $N_f=3$ flavor case. Numerical 
results are shown in Fig.~\ref{fig_gap_inst}. We observe
that the instanton correction becomes large for $\mu
\sim 1250$ MeV in $N_f=2$ QCD and for $\mu\sim 750$ MeV
in the realistic case of QCD with three flavors and
a massive strange quark. 

\begin{figure}
\begin{center}
\leavevmode
\includegraphics[width=7cm,clip=true]{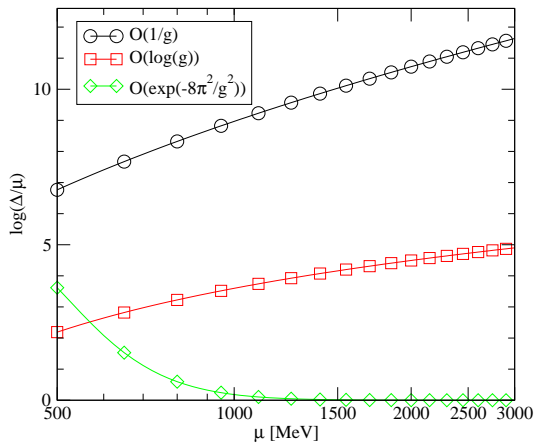}
\end{center}
\caption{This figure shows successive terms in the 
perturbative expansion of $\log(\mu/\Delta)$ in 
QCD with $N_f=3$ flavors.}
\label{fig_pert}
\end{figure}

 We should note that the fact that the instanton correction 
to the gap is on the order of 100\% does not necessarily
invalidate the perturbative expansion. As explained in 
the previous section, the quantity that is being expanded
is not the gap $\Delta$, but $\log(\mu/\Delta)$. In 
Fig.~\ref{fig_pert} we compare the size of the leading 
order $O(1/g)$ term, the sub-leading $O(\log(g))$ and $O(1)$ 
terms, as well as the instanton contribution to the logarithm
of the gap in $N_f=3$ QCD. We observe that exponentially
small instanton contributions start to dominate over all 
other contact terms for $\mu\sim 600$ MeV. 

 For $\mu\sim 500$ MeV the instanton term becomes comparable 
to the leading $O(1/g)$ term and we can no longer consider
instantons effects to be a small correction. In this regime
it probably makes more sense to do an instanton calculation
and consider perturbative gluon exchanges as a correction. This 
is the approach originally suggested in \cite{Rapp:1998zu,Alford:1999mk}. 
We should note, however, that for $\mu<500$ MeV the instanton
calculation requires some phenomenological input because
the instanton size distribution equ.~(\ref{G_I}) is no 
longer reliably calculable. 
 
 Ideally, we would like to check perturbative calculations
of the gap against numerical calculations on the lattice. 
While this cannot be done in the realistic case of QCD with
$N_c=3$ colors, the comparison is possible for QCD with 
$N_c=2$ colors and an even number of flavors, or for QCD
with $N_c=3$ colors and a non-zero chemical potential 
for isospin rather than baryon number. In both of these
cases it is also possible to separate the perturbative and 
instanton contributions by measuring both the gap and the mass 
of the $U(1)_A$ Goldstone boson \cite{Schafer:2002yy}. 

\begin{figure}
\begin{center}
\leavevmode
\includegraphics[width=7cm,clip=true]{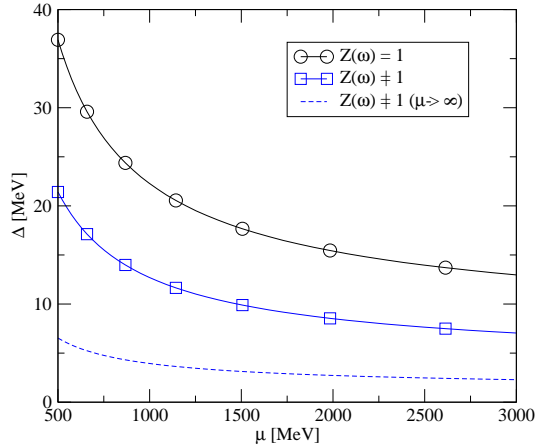}
\end{center}
\caption{Numerical solution of the perturbative gap 
equation. We show the results with and without the 
fermion self energy included. We also show the 
asymptotic result, which corresponds to a reduction 
of the gap by a factor $\exp(-(\pi^2+4)/8)\simeq 0.18$.}
\label{fig_gap_z}
\end{figure}

\section{Numerical Estimates} 
\label{sec_num}

 In this section we shall address a few more issues 
related to the magnitude of the gap. We will estimate 
the size of certain higher order corrections using 
numerical solutions of the gap equation.
The first problem we wish to study is the importance
of the fermion self energy correction. In Sect.~\ref{sec_cor}
we saw that at asymptotically large chemical potential 
non-Fermi liquid effects in the fermion self energy 
reduce the gap by a factor $\exp(-(\pi^2+4)/8)\simeq 0.18$. 
In order to study the problem at moderate densities we consider 
the gap equation
\be 
\label{gap_z}
\Delta(p_4) =\frac{g^2}{18\pi^2}
\int_0^{\mu} dq_4\, Z(q_4)
\frac{\Delta(q_4)}{\sqrt{q_4^2+\Delta(q_4)^2}}
\log\left( \frac{b_0 \mu}{|p_4^2-q_4^2|^{1/2}}\right),
\ee
with 
\be
\label{Z}
Z(q_4)=\left(1+\frac{g^2}{9\pi^2}\log
 \left(\frac{m_D}{q_4}\right)\right)^{-1}
\ee
and $b_0=256\pi^4g^{-5}$. If the density is very
large then the scale inside the logarithm in 
equ.~(\ref{Z}) does not matter. For our numerical 
estimates we have used the electric screening mass
which is the scale suggested by one-loop calculations
\cite{Manuel:2000mk}. 

\begin{figure}
\begin{center}
\leavevmode
\includegraphics[width=7cm,clip=true]{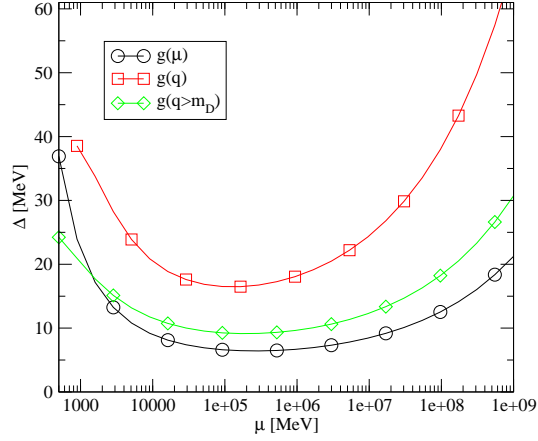}
\end{center}
\caption{Numerical solution of the gap equation with
running coupling constant effects included. We show 
the result with a fixed coupling $g=g(\mu)$, with 
a coupling that runs as a function of the momentum
transfer as long as the momentum is larger than 
the electric screening scale, and a coupling that 
runs as a function of the momentum without any 
restrictions.}
\label{fig_gap_rg}
\end{figure}

 Numerical results are shown in Fig.~\ref{fig_gap_z}.
We observe that the gap is indeed reduced by the effects
of fermion wave function renormalization, but at moderate 
density the reduction is smaller as compared to the asymptotic
result. This is related to the fact that for $\mu\sim 500$
MeV and $\Delta\sim 50$ MeV the logarithm $\log(m_D/
\Delta)$, which is formally $O(1/g)$, is numerically 
not very large.

 The second issue we wish to address is the role 
of running coupling constant effects in the gap 
equation \cite{Beane:2000hu}. For the numerical 
estimates presented in the last section we have 
used the weak coupling result equ.~(\ref{gap_og})
with the coupling constant evaluated at the scale
$\mu$. It is easy to see that variation in the 
scale correspond to $O(g^2)$ corrections, which 
is of the same order as other terms that have 
been neglected. 

 In order to estimate the size of these effects 
we consider the gap equation
\bea 
\label{gap_rg}
\Delta(p_4) &=&
\int_0^{\mu} dq_4\, 
 \bigg\{ \frac{g^2(m_D)}{18\pi^2}
   \log\left(\frac{m_D^3+\Lambda_M^3}{\Lambda_M^3}
   \right) 
 + 2\log\bigg(\left[ \frac{g^2(2\mu)}{g^2(m_D)}
    \right]^{4/(3\beta_0)}\bigg) \bigg\} \nonumber \\
 & & \mbox{}\hspace{1cm} \cdot Z(q_4)
\frac{\Delta(q_4)}{\sqrt{q_4^2+\Delta(q_4)^2}} ,
\eea
where $\Lambda_M=(\pi m_D^2 |p_4\pm q_4|/4)^{1/3}$
is the scale that characterizes magnetic gluon exchanges. 
For momenta above the electric screening scale we have used 
the one-loop running coupling at the scale set by the momentum 
transfer. For momenta below the screening scale the coupling is 
frozen. In the language if the high density effective theory 
this means that the we have performed the matching at the electric 
screening scale. The four-fermion operator acquires an anomalous 
dimension which is equal to the QCD beta function \cite{Schafer:xx}. 
Below the electric screening scale the gluonic interaction is 
effectively three-dimensional and does not run. 

 In Fig.~\ref{fig_gap_rg} we show results obtained by solving 
equ.~(\ref{gap_rg}) numerically. For comparison we also show 
results obtained by solving a gap equation in which the coupling 
is allowed to run all the way down to the magnetic scale $(m_D^2
\Delta)^{1/3}$. This approximation was suggested by Beane et 
al.~\cite{Beane:2000hu}. We observe that for a moderate 
chemical potential $\mu\sim 500$ MeV the running coupling 
constant does not lead to very large effects. The reason is 
that there is no large hierarchy between the scales $m_D$ and
$\mu$. If the chemical potential is very large, $\mu\sim
10$ GeV, the gap is increased by about 50\%. This effect
slowly disappears at asymptotically large chemical potential. 
The situation is different in case of the gap equation proposed 
by Beane et al. In that case the gap equation involves an 
extra large logarithm $\log(\mu/\Delta)$ and the pre-exponential
factor in the asymptotic solution is modified \cite{Beane:2000hu}.

\section{Conclusions} 
\label{sec_con}

 We discussed an effective field theory for QCD at high 
baryon density. We studied, in particular, the problem of 
power counting in the high density effective theory. We 
showed that the power counting is complicated by ``hard 
dense loops'', i.e. loop diagrams that involve the large 
scale $\mu^2$ and proposed a power counting that takes these
effects into account. The modified $l$ counting implies
that hard dense loops in gluon $n$-point functions have
to be resummed below the scale $g\mu$, and that four
fermion operators are leading order in the HDET power
counting. 

 We used the high density effective theory to study the 
size of instanton corrections to the gap in superfluid
quark matter. We found that instanton effects are very 
large in the regime $\mu\sim 500$ MeV which is of physical 
interest. We argued that numerical calculation in QCD 
with $N_c=2$ colors, or QCD with $N_c=3$ colors and 
non-zero isospin chemical potential, will of great 
help in determining the gap in $N_c=3$ QCD at non-zero
baryon density. We also studied a renormalization group 
improved gap equation. We find no significant corrections
as compared to a gap equation with a fixed coupling. 
 
 Acknowledgments: This work was supported in part by US DOE 
grant DE-FG-88ER40388.

\end{document}